# Concurrent Models for Object Execution

*Bob Diertens*

section Theory of Computer Science, Faculty of Science, University of Amsterdam

*ABSTRACT*

In previous work we developed a framework of computational models for the concurrent execution of functions on different levels of abstraction. It shows that the traditional sequential execution of function is just a possible implementation of an abstract computational model that allows for the concurrent execution of functions. We use this framework as base for the development of abstract computational models that allow for the concurrent execution of objects.

## 1. Introduction

To execute a program written in a particular programming language, it is compiled into executable code for a particular machine. The machine is actually a machine model representing physical hardware, operation system, etc, or possibly a virtual machine. The compilation is done according to an execution model specific for the machine model. An execution model is an implementation of a computational model[1] which gives the essential rules for performing computations. The computational model must at least be adequate for expressing the operational semantics of the programming language. An overview of this all is given in Figure 1. For long the machine model was based on sequential execution of instructions. Programming languages were based on sequential execution of instructions as well, as were the computational models and the execution models.

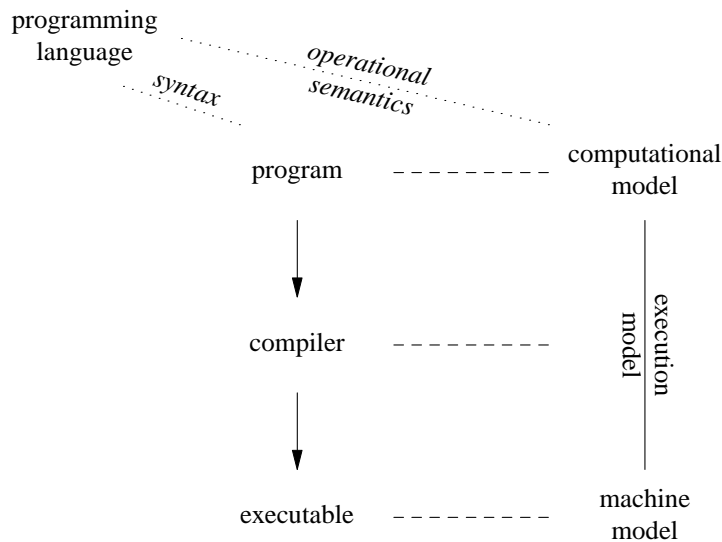

**Figure 1.** Overview of program execution related affairs

---

1.   A computational model is also called an abstract machine model, although the terms can be considered different elsewhere.

In previous work [6] we developed a framework of computational models for the concurrent execution of functions. This framework can be used for the development of concurrent computational models that deal with the problems inherent with concurrency. It also shows that it should not be decided which parts of a sequential system can be done concurrently, but what parts of a concurrent system can or should be done sequentially. In this article we set out to do the same for the concurrent execution of objects. Models for object execution make use of models for function execution. The invocation of a method for a particular function is implemented as the sequential execution of a function. Concurrency of objects is added in the same way as concurrency is added to functions. Either this is done using an add-on library, such as multi-threading [3], or through extension of the programming languages with constructs for concurrency.

These extensions, however, are implemented on top of existing computational and execution models using the same add-on libraries. The problem with adding concurrency in this way is that it is still based on a computational model for sequential execution of instructions, and compilation is still based on a model for sequential execution of instructions. Compilers may generate efficient code that is correct for sequential execution, but incorrect for concurrent execution. This was already shown in [5] (1995) and again later in [4] (2005) and is caused by communications between threads through shared memory. In both [5] and [4] it is stated that concurrency must be addressed at the language specification level and in compiler design.

There are of course also new programming languages (or redesigned existing ones) which support concurrency that come with computational models and execution models which solve some of the problems, if not all, of concurrency, such as Java, C#, and Ada. But it is quite tricky to avoid problems with memory models for such a language as is shown in [8] and [7] for Java, and in [1] and [2] for a C++ standard.

In section 2 we summarize our previous work on concurrent models for function execution. We extend this work with concurrent models for object execution in section 3.

## 2. Computational models for functions

In [6] we developed a framework of computational models at different levels of abstraction for the concurrent execution of functions. The development of the framework started with the traditional sequential execution model for functions from which a sequential computational model was obtained by abstracting from the details of function call implementation. Further abstraction from the way a function is scheduled for execution led to an abstract computational model that allows for the concurrent execution of functions (Figure 2).

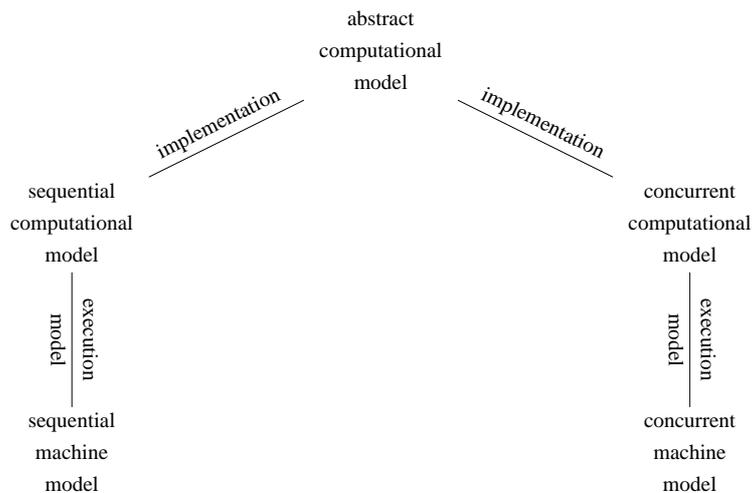

**Figure 2.** Framework of computational models

This approach shows that with abstraction and relaxing constraints, a model for execution of functions can be obtained in which function scheduling plays a key role. This model has as a possible implementation inline scheduling, the original stack-based function execution model the development of the framework started with. But more important, this model allows for the concurrent execution of functions, and therefore it can be used as a model for the implementation of concurrent software. The framework of computational models at different levels of abstraction can be used for further development of concurrent computational models that deal with the problems inherent with concurrency.

## 3. Computational models for objects

The traditional sequential execution model for objects is based on the traditional sequential execution model for functions. In this model it seems that the objects are not executed, but the methods of the objects are executed, and in a similar manner as functions. We could therefore use the same framework of computational models for the methods of objects as the one developed for functions (Figure 2). This framework then allows for the concurrent executions of methods.

We can extend the framework so that it allows for the concurrent execution of objects. For this we have to see that the scheduler for the methods has to make a context switch to the object a method belongs to. In this framework the context switch is done inline with the scheduling of a method, similar to the inline scheduling of functions in the sequential computational model for functions. Apparently, the context switch is the scheduling of the object. Separating the scheduling of objects from the scheduling of methods gives us a scheduler for objects where each object takes care of the scheduling of its methods.

With this, we abstracted from how objects are scheduled. The result is a computational model that allows for the concurrent execution of objects. In this model, each object decides how to schedule its methods. Of course, the scheduling of methods can be inline and thus limiting concurrency to objects only. From this model we can implement the abstract computational model that allows for the concurrent execution of functions or methods, but we can also implement a model that allows for the concurrent execution of objects in which each object decides how to schedule its methods.

What has been a call of a method for a particular object in the sequential execution model has now become a request to that object. How the object handles the request is not important, and so we can abstract from this.

## 4. Conclusions

In this article we extended the framework of computational models for function execution into a framework of computational models for object execution. This framework allows for the concurrent execution of objects where each object controls the scheduling of its own methods. This framework can be used as model for the implementation of software where the software may be object-oriented or function-oriented. The type of scheduling in the implementation of the system or part of the system at hand can be chosen such that there can be made full use of the capabilities of the underlying machine model. The main advantage is that the problems inherent with concurrency can be dealt with at the right level of abstraction.